\documentclass[sigconf]{acmart} 

\copyrightyear{2024}
\acmYear{2024}
\setcopyright{rightsretained}
\acmConference[RecSys '24]{18th ACM Conference on Recommender Systems}{October 14--18, 2024}{Bari, Italy}
\acmBooktitle{18th ACM Conference on Recommender Systems (RecSys '24), October 14--18, 2024, Bari, Italy}\acmDOI{10.1145/3640457.3688059}
\acmISBN{979-8-4007-0505-2/24/10}

\usepackage{bm}
\usepackage{bbm}

\begin{document}

\title[Multinomial Blending]{Ranking Across Different Content Types:\\The Robust Beauty of Multinomial Blending}

\author{Jan Malte Lichtenberg}
\authornote{Both authors contributed equally to this research.}
\email{jlichten@amazon.de}
\affiliation{%
  \institution{Amazon Music}
  \country{Germany}
}
\author{Giuseppe Di Benedetto}
\authornotemark[1]
\email{bgiusep@amazon.de}
\affiliation{%
  \institution{Amazon Music}
  \country{Germany}
}
\author{Matteo Ruffini}
\authornote{Work done while working at Amazon Music.}
\email{matteo@usealbatross.ai}
\affiliation{%
  \institution{Albatross AI}
  \country{Germany}
}

\renewcommand{\shortauthors}{Lichtenberg, Di Benedetto, Ruffini}

\begin{abstract}
An increasing number of media streaming services have expanded their offerings to include entities of multiple content types. For instance, audio streaming services that started by offering music only, now also offer podcasts, merchandise items, and videos. Ranking items across different content types into a single slate poses a significant challenge for traditional learning-to-rank (LTR) algorithms due to differing user engagement patterns for different content types. We explore a simple method for cross-content-type ranking, called multinomial blending (MB), which can be used in conjunction with most existing LTR algorithms. We compare MB to existing baselines not only in terms of ranking quality but also from other industry-relevant perspectives such as interpretability, ease-of-use, and stability in dynamic environments with changing user behavior and ranking model retraining. Finally, we report the results of an A/B test from an Amazon Music ranking use-case. 
\end{abstract}





\maketitle

\section{Introduction}
It is increasingly common for streaming media services to offer content of different types the users can engage with. For instance, news websites often show articles and videos; music streaming platforms can provide music, podcasts, videos, and merchandise.
Traditionally, many of these main home pages follow a retrieve and re-ranking approach. Whereas the retrieval step can just be turned into multiple retrieval steps (e.g., one per content type), applying traditional learning-to-rank (LTR) algorithms to the multi-content-type setting comes with several challenges: 1) disjoint item feature sets across content types; 2) cold-start problem if one content type is introduced in a later stage; and 3) different engagement patterns and reward signals. For instance, users tend to listen to music tracks much more frequently and repeatedly than podcast episodes, which are typically listened to only once and less frequently due to their longer time commitment.

These discrepancies can lead to highly unbalanced exposure for items of specific content types. In particular, if short-term reward signals such as clicks are used to optimize the ranking policy, items from low-frequency engagement (or \textit{slow}) content types (e.g., podcasts) might be disadvantaged due to their low average engagement, despite potentially being more relevant or having higher long-term value than items from high-frequency engagement (or \textit{fast}) content types (e.g., music). 
Such reduced level of exposure can lead to even less engagement, which in turn leads the ranking algorithm to provide even less exposure to the slow items. This vicious circle maximizes short-term engagement but drowns out items from slow content types, which can ultimately lead to decreased long-term user satisfaction and negative effects on content creators from the slower content types.

In this work, we analyze the cross-content-type ranking problem using the example of ranking podcasts and music widgets on a single page. The diversity and fairness literature~\cite[e.g.,][]{singh2018fairness, li2020cascading, saito2022fair, oosterhuis2021computationally} provides various approaches that could be adapted to the cross-content-type ranking problem. However, as described below, these methods, which were often evaluated in academic settings with static data sets, are difficult to employ in dynamic, industry-scale settings.
 We propose a simple-yet-efficient method called \textit{multinomial blending} (MB) which trades-off personalized diversification in favour of interpretability and ease-of-use in order to comply with business requirements described in Section \ref{CCT}. In Section \ref{experiments} we describe how we use MB in practice at Amazon Music and report the results of an A/B test.

\section{The Cross-Content-Type (CCT) ranking problem} \label{CCT}

\textbf{Setting.} We are interested in ranking a set of $n$ items $d_{j}, j = 1, \dots, n$, where each item belongs to a single content type $c(d_j) \in \{1, \dots, C\}$, into a slate of length $k$. Score-based learning-to-rank (LTR) algorithms~\citep{haldar2020improving} learn a function $h: \mathbb{R}^p \rightarrow \mathbb{R}$ that scores each item using a feature representation $\bm{x}_j\in \mathbb{R}^p$ as input~\citep{mcinerney2018explore}. The ranking is then typically produced either by deterministically sorting the items by scores or via repeated softmax sampling~\citep{singh2018fairness}. We assume that the score-based LTR algorithm is frequently re-trained to tackle non-stationarity.

\noindent \textbf{Requirements.} 
The overarching goal for the CCT ranking model is to provide a Pareto improvement in some engagement metric (for example, click-through rate). More specifically, the ranker should boost engagement with items from the slow content type while not harming overall engagement. In an industry setting, we additionally define the following product requirements for the CCT ranker:  
\begin{itemize}
    \item controllable content-type exposure budgets through interpretable parameters;
    \item compatibility with model retraining and stability in non-stationary user environments;  
    \item personalized ranking within content-types.
\end{itemize}
In our use-case, exposure budgets for different content types are product requirements that are defined at business level. Such requirements are not considered by classic LTR algorithms optimizing for a short term metric such as clicks. Therefore different approaches have been proposed with the goal of boosting exposure of specific content types.

\noindent \textbf{Policies that optimize for diversity.} To tackle the content-type diversity problem, some works have tried to optimize for diversity by changing the loss function of the LTR algorithm~\citep{singh2018fairness, li2020cascading, saito2022fair, oosterhuis2021computationally}. Despite being formally sound, such approaches are not easy to apply in practice: (i) they lack content-type exposure guarantees; (ii) the hyper-parameters regulating the trade-off between ranking performance and diversity are difficult to interpret; and (iii) model retraining usually requires re-tuning of the diversity parameter to ensure that content-type exposure remains as desired.

\noindent \textbf{Reward shaping}~\citep{ng1999policy}. Another option is to assign different weights to clicks from different content types. However, estimating the relative importance of reward signals across various content types is a difficult endeavour on its own, and predicting how this reward shaping will influence the actual exposure of different content types becomes even more complex, particularly in environments characterized by  dynamic user behavior patterns.

\noindent \textbf{Post-processing approaches to ensuring diversity.} One way of ensuring that items from each content type receive sufficient exposure is to put in place manual \textit{overrides} that tie particular items from under-represented content types to certain positions. While this strategy can indeed provide exposure to slow content types, it has several down-sides: the manual overrides are slightly personalized or not personalized at all; overrides need to be curated and lifecycle-managed requiring additional labor; and they can bias propensity estimates for counterfactual off-policy estimators~\citep{jakimov2023unbiased, li2018offline, di2023contextual, buchholz2024counterfactual}.
More principled post-processing approaches include various re-ranking methods, which perturb the initial, personalized ranking to satisfy diversity constraints while still retaining personalization information contained in the ranking scores. For instance, the \textit{maximal marginal relevance} (MMR)~\citep{carbonell1998use} approach to produces diverse rankings can be easily adapted to the CCT ranking problem (see Appendix). That said, the MMR approach suffers from the same limitations outlined further above for diversity-optimizing policies.

In the next section, we discuss a simple alternative approach that directly reasons about the exposure given to the various content types as opposed to approaches that control exposure indirectly via modifications of reward signals or loss functions.

\section{Multinomial Blending (MB)}

\begin{figure}
    \centering
    \includegraphics[width=0.499\textwidth]{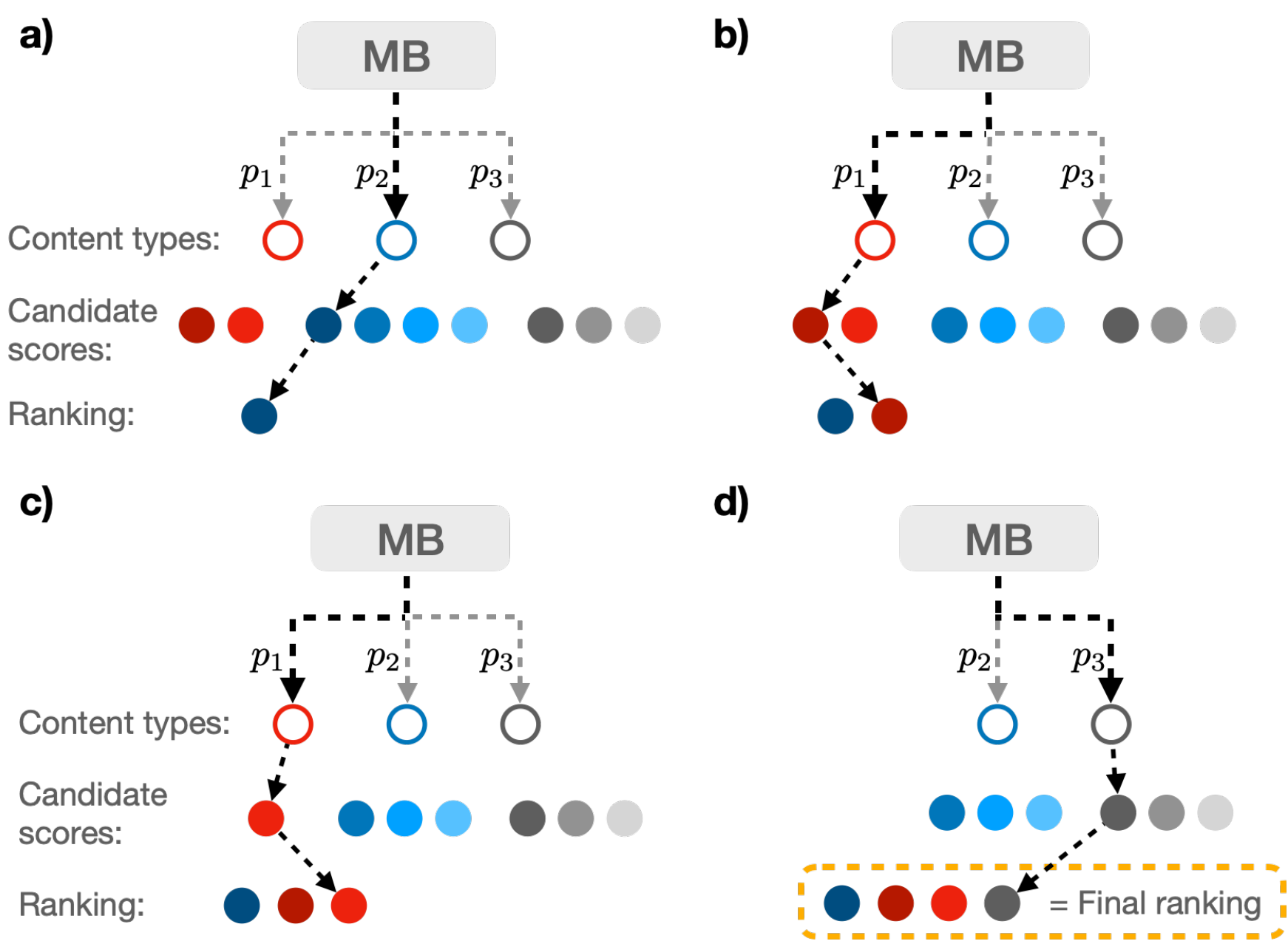}
    \Description[Illustration of multinomial blending.]{A toy illustration of multinomial blending for three content types.}
    \caption{Multinomial blending (MB) for $C=3$ content types (red, blue, gray) and a ranking size of $k=4$. At each step, the blender samples a content type according to sampling probabilities $\bm{p} = [p_1, p_2, p_3]$ and then selects the highest-scoring remaining candidate. In this example, the candidate pool for the red content type is empty after the third iteration (sub-figure \textbf{c}), leading to re-normalization of the sampling probabilities for remaining content types.  
    }
    \label{fig:mb_illustration}
\end{figure}

Multinomial blending is defined by a multinomial probability distribution $\mathcal{M}$ with vector of sampling probabilities $\bm{p} = [p_1, p_2, \dots, p_C]$, potentially defined by the business team, where $C$ is the number of content types. Figure \ref{fig:mb_illustration} illustrates the blending procedure. The scoring function $h$ is used once to score all candidate items (darker color indicates higher score) and candidates are ranked by decreasing scores within each content type. To select the next action, MB first samples a content type according to $c \sim \mathcal{M}(\bm{p})$ and then selects the highest-scoring remaining candidate from that content type. This procedure is repeated until the slate is filled to the desired size ($k=4$ in Figure \ref{fig:mb_illustration}).
MB satisfies the desiderata listed above:
\begin{itemize}
    \item Interpretability. Each parameter $p_c$ maps to the expected content-type average exposure: on average, $p_c * 100$\% of the slate will be covered by items from content type $c$. 
    \item Stability. The average exposure guarantees are independent of the underlying scoring function $h$ and therefore remain stable even after model re-training or non-stationary user behavior.
    \item Within-content-type personalization. The personalized ranking of items within each content type is preserved, thereby preserving personalization quality as learned by the original scoring function.
\end{itemize}
One downside of MB is that content-type exposure itself is \textit{not} personalized, that is, average exposure allocations of the various content types are constant across different users and prediction contexts. Developing methods that achieve full personalization while maintaining the stability and interpretability benefits of MB is an interesting direction for future work. In the Appendix, we describe a heuristic modification of MB that replaces the equality exposure guarantees by lower-bound exposure guarantees.

\section{Experiments} \label{experiments}

We A/B-tested multinomial blending on an Amazon Music CCT ranking problem where podcast and music containers were ranked into the same slate. To counter-act under-exposure as produced by the production ranker (a Deep PropDCG model~\citep{agarwal2019general}), the existing approach (control treatment) was to define a set of manual ranking overrides that would boost podcast exposure in a slightly personalized manner. We compared the existing approach to a) using the existing ranker with MMR diversification and b) using the existing ranker with MB diversification. For the MMR treatment, the diversity parameter was tuned offline to ensure the desired podcast exposure. During offline evaluation, operational limitations of the MMR approach became apparent: various market places required different MMR penalty parameters and optimal diversification rates would change over time. On the other hand, for the MB approach, we were able to select a single diversification rate, reducing the operational burden of having to maintain different model versions. Furthermore, we observed that restarting the model could lead to changing score distribution and thus to different diversification behavior. Table \ref{tab:weblab_results} shows that both treatments obtained a Pareto-improvement over the control treatment. 
\begin{table}[h] 
\centering
\begin{tabular}{lcc}
\toprule
\textbf{Algorithm} & \textbf{Podcast listening time} & \textbf{Overall engagement} \\
\midrule
MMR & +13.57\% & \textbf{+2.76\%} \\
MB & \textbf{+18.82\%} & +2.23\% \\
\bottomrule
\end{tabular}
\caption{Lift in podcast listening time and overall engagement compared to the control treatment.}
\label{tab:weblab_results}
\end{table}

\section{Discussion and Conclusion}
Despite in the A/B test both MMR and MB achieved a Pareto improvement in podcast listening time and the overall engagement metric, MB was globally launched due to its operational advantages. 
Other than the launch metrics, MB outperformed MMR in podcast user acquisition metrics (e.g., count of users' first podcast streams), which is expected as MMR tends to surface fewer podcasts to users who are not interested in this content type, while MB ensures a certain podcast exposure at user level.
As discussed above, MB is easy and intuitive to set up and update over time (e.g., if business goals change), while MMR requires counterfactual evaluation (and hence data collection) in order for its trade-off parameter to be tuned. Moreover, MB is computationally more efficient, as it does not require sequential re-computation of the scores as in MMR (see Appendix), and it allows to compute the propensity matrix in closed form, which can be used in off-policy evaluation \citep{li2018offline} (see Appendix).

\bibliographystyle{ACM-Reference-Format}
\bibliography{sample-base}


\begin{thebibliography}{14}


\ifx \showCODEN    \undefined \def \showCODEN     #1{\unskip}     \fi
\ifx \showDOI      \undefined \def \showDOI       #1{#1}\fi
\ifx \showISBNx    \undefined \def \showISBNx     #1{\unskip}     \fi
\ifx \showISBNxiii \undefined \def \showISBNxiii  #1{\unskip}     \fi
\ifx \showISSN     \undefined \def \showISSN      #1{\unskip}     \fi
\ifx \showLCCN     \undefined \def \showLCCN      #1{\unskip}     \fi
\ifx \shownote     \undefined \def \shownote      #1{#1}          \fi
\ifx \showarticletitle \undefined \def \showarticletitle #1{#1}   \fi
\ifx \showURL      \undefined \def \showURL       {\relax}        \fi
\providecommand\bibfield[2]{#2}
\providecommand\bibinfo[2]{#2}
\providecommand\natexlab[1]{#1}
\providecommand\showeprint[2][]{arXiv:#2}

\bibitem[Agarwal et~al\mbox{.}(2019)]%
        {agarwal2019general}
\bibfield{author}{\bibinfo{person}{Aman Agarwal}, \bibinfo{person}{Kenta
  Takatsu}, \bibinfo{person}{Ivan Zaitsev}, {and} \bibinfo{person}{Thorsten
  Joachims}.} \bibinfo{year}{2019}\natexlab{}.
\newblock \showarticletitle{A general framework for counterfactual
  learning-to-rank}. In \bibinfo{booktitle}{\emph{Proceedings of the 42nd
  International ACM SIGIR Conference on Research and Development in Information
  Retrieval}}. \bibinfo{pages}{5--14}.
\newblock


\bibitem[Buchholz et~al\mbox{.}(2024)]%
        {buchholz2024counterfactual}
\bibfield{author}{\bibinfo{person}{Alexander Buchholz}, \bibinfo{person}{Ben
  London}, \bibinfo{person}{Giuseppe Di~Benedetto}, \bibinfo{person}{Jan~Malte
  Lichtenberg}, \bibinfo{person}{Yannik Stein}, {and} \bibinfo{person}{Thorsten
  Joachims}.} \bibinfo{year}{2024}\natexlab{}.
\newblock \showarticletitle{Counterfactual ranking evaluation with flexible
  click models}. In \bibinfo{booktitle}{\emph{Proceedings of the 47th
  International ACM SIGIR Conference on Research and Development in Information
  Retrieval}}. \bibinfo{pages}{1200--1210}.
\newblock


\bibitem[Carbonell and Goldstein(1998)]%
        {carbonell1998use}
\bibfield{author}{\bibinfo{person}{Jaime Carbonell} {and} \bibinfo{person}{Jade
  Goldstein}.} \bibinfo{year}{1998}\natexlab{}.
\newblock \showarticletitle{The use of MMR, diversity-based reranking for
  reordering documents and producing summaries}. In
  \bibinfo{booktitle}{\emph{Proceedings of the 21st annual international ACM
  SIGIR conference on Research and development in information retrieval}}.
  \bibinfo{pages}{335--336}.
\newblock


\bibitem[Di~Benedetto et~al\mbox{.}(2023)]%
        {di2023contextual}
\bibfield{author}{\bibinfo{person}{Giuseppe Di~Benedetto},
  \bibinfo{person}{Alexander Buchholz}, \bibinfo{person}{Ben London},
  \bibinfo{person}{Matej Jakimov}, \bibinfo{person}{Yannik Stein},
  \bibinfo{person}{Jan~Malte Lichtenberg}, \bibinfo{person}{Vito Bellini},
  {and} \bibinfo{person}{Matteo Ruffini}.} \bibinfo{year}{2023}\natexlab{}.
\newblock \showarticletitle{Contextual position bias estimation using a single
  stochastic logging policy}.
\newblock  (\bibinfo{year}{2023}).
\newblock


\bibitem[Haldar et~al\mbox{.}(2020)]%
        {haldar2020improving}
\bibfield{author}{\bibinfo{person}{Malay Haldar}, \bibinfo{person}{Prashant
  Ramanathan}, \bibinfo{person}{Tyler Sax}, \bibinfo{person}{Mustafa Abdool},
  \bibinfo{person}{Lanbo Zhang}, \bibinfo{person}{Aamir Mansawala},
  \bibinfo{person}{Shulin Yang}, \bibinfo{person}{Bradley Turnbull}, {and}
  \bibinfo{person}{Junshuo Liao}.} \bibinfo{year}{2020}\natexlab{}.
\newblock \showarticletitle{Improving deep learning for airbnb search}. In
  \bibinfo{booktitle}{\emph{Proceedings of the 26th ACM SIGKDD International
  Conference on Knowledge Discovery \& Data Mining}}.
  \bibinfo{pages}{2822--2830}.
\newblock


\bibitem[Jakimov et~al\mbox{.}(2023)]%
        {jakimov2023unbiased}
\bibfield{author}{\bibinfo{person}{Matej Jakimov}, \bibinfo{person}{Alexander
  Buchholz}, \bibinfo{person}{Yannik Stein}, {and} \bibinfo{person}{Thorsten
  Joachims}.} \bibinfo{year}{2023}\natexlab{}.
\newblock \showarticletitle{Unbiased offline evaluation for learning to rank
  with business rules}.
\newblock  (\bibinfo{year}{2023}).
\newblock


\bibitem[Li et~al\mbox{.}(2020)]%
        {li2020cascading}
\bibfield{author}{\bibinfo{person}{Chang Li}, \bibinfo{person}{Haoyun Feng},
  {and} \bibinfo{person}{Maarten~de Rijke}.} \bibinfo{year}{2020}\natexlab{}.
\newblock \showarticletitle{Cascading hybrid bandits: Online learning to rank
  for relevance and diversity}. In \bibinfo{booktitle}{\emph{Proceedings of the
  14th ACM Conference on Recommender Systems}}. \bibinfo{pages}{33--42}.
\newblock


\bibitem[Li et~al\mbox{.}(2018)]%
        {li2018offline}
\bibfield{author}{\bibinfo{person}{Shuai Li}, \bibinfo{person}{Yasin
  Abbasi-Yadkori}, \bibinfo{person}{Branislav Kveton}, \bibinfo{person}{Shan
  Muthukrishnan}, \bibinfo{person}{Vishwa Vinay}, {and} \bibinfo{person}{Zheng
  Wen}.} \bibinfo{year}{2018}\natexlab{}.
\newblock \showarticletitle{Offline evaluation of ranking policies with click
  models}. In \bibinfo{booktitle}{\emph{Proceedings of the 24th ACM SIGKDD
  International Conference on Knowledge Discovery \& Data Mining}}.
  \bibinfo{pages}{1685--1694}.
\newblock


\bibitem[McInerney et~al\mbox{.}(2018)]%
        {mcinerney2018explore}
\bibfield{author}{\bibinfo{person}{James McInerney}, \bibinfo{person}{Benjamin
  Lacker}, \bibinfo{person}{Samantha Hansen}, \bibinfo{person}{Karl Higley},
  \bibinfo{person}{Hugues Bouchard}, \bibinfo{person}{Alois Gruson}, {and}
  \bibinfo{person}{Rishabh Mehrotra}.} \bibinfo{year}{2018}\natexlab{}.
\newblock \showarticletitle{Explore, exploit, and explain: personalizing
  explainable recommendations with bandits}. In
  \bibinfo{booktitle}{\emph{Proceedings of the 12th ACM conference on
  recommender systems}}. \bibinfo{pages}{31--39}.
\newblock


\bibitem[Ng et~al\mbox{.}(1999)]%
        {ng1999policy}
\bibfield{author}{\bibinfo{person}{Andrew~Y Ng}, \bibinfo{person}{Daishi
  Harada}, {and} \bibinfo{person}{Stuart Russell}.}
  \bibinfo{year}{1999}\natexlab{}.
\newblock \showarticletitle{Policy invariance under reward transformations:
  Theory and application to reward shaping}. In
  \bibinfo{booktitle}{\emph{Icml}}, Vol.~\bibinfo{volume}{99}.
  \bibinfo{pages}{278--287}.
\newblock


\bibitem[Oosterhuis(2021)]%
        {oosterhuis2021computationally}
\bibfield{author}{\bibinfo{person}{Harrie Oosterhuis}.}
  \bibinfo{year}{2021}\natexlab{}.
\newblock \showarticletitle{Computationally efficient optimization of
  plackett-luce ranking models for relevance and fairness}. In
  \bibinfo{booktitle}{\emph{Proceedings of the 44th International ACM SIGIR
  Conference on Research and Development in Information Retrieval}}.
  \bibinfo{pages}{1023--1032}.
\newblock


\bibitem[Saito and Joachims(2022)]%
        {saito2022fair}
\bibfield{author}{\bibinfo{person}{Yuta Saito} {and} \bibinfo{person}{Thorsten
  Joachims}.} \bibinfo{year}{2022}\natexlab{}.
\newblock \showarticletitle{Fair Ranking as Fair Division: Impact-Based
  Individual Fairness in Ranking}. In \bibinfo{booktitle}{\emph{Proceedings of
  the 28th ACM SIGKDD Conference on Knowledge Discovery and Data Mining}}.
  \bibinfo{pages}{1514--1524}.
\newblock


\bibitem[Singh and Joachims(2018)]%
        {singh2018fairness}
\bibfield{author}{\bibinfo{person}{Ashudeep Singh} {and}
  \bibinfo{person}{Thorsten Joachims}.} \bibinfo{year}{2018}\natexlab{}.
\newblock \showarticletitle{Fairness of exposure in rankings}. In
  \bibinfo{booktitle}{\emph{Proceedings of the 24th ACM SIGKDD international
  conference on knowledge discovery \& data mining}}.
  \bibinfo{pages}{2219--2228}.
\newblock


\bibitem[Virtanen et~al\mbox{.}(2020)]%
        {virtanen2020scipy}
\bibfield{author}{\bibinfo{person}{Pauli Virtanen}, \bibinfo{person}{Ralf
  Gommers}, \bibinfo{person}{Travis~E Oliphant}, \bibinfo{person}{Matt
  Haberland}, \bibinfo{person}{Tyler Reddy}, \bibinfo{person}{David
  Cournapeau}, \bibinfo{person}{Evgeni Burovski}, \bibinfo{person}{Pearu
  Peterson}, \bibinfo{person}{Warren Weckesser}, \bibinfo{person}{Jonathan
  Bright}, {et~al\mbox{.}}} \bibinfo{year}{2020}\natexlab{}.
\newblock \showarticletitle{SciPy 1.0: fundamental algorithms for scientific
  computing in Python}.
\newblock \bibinfo{journal}{\emph{Nature methods}} \bibinfo{volume}{17},
  \bibinfo{number}{3} (\bibinfo{year}{2020}), \bibinfo{pages}{261--272}.
\newblock


\end{thebibliography}




\newpage

\appendix

\section{Appendix}

\subsection{MMR selection for ranking diversification.} \label{MMR}
\begin{figure*}[h]
    \centering
    \includegraphics[width=0.98\textwidth]{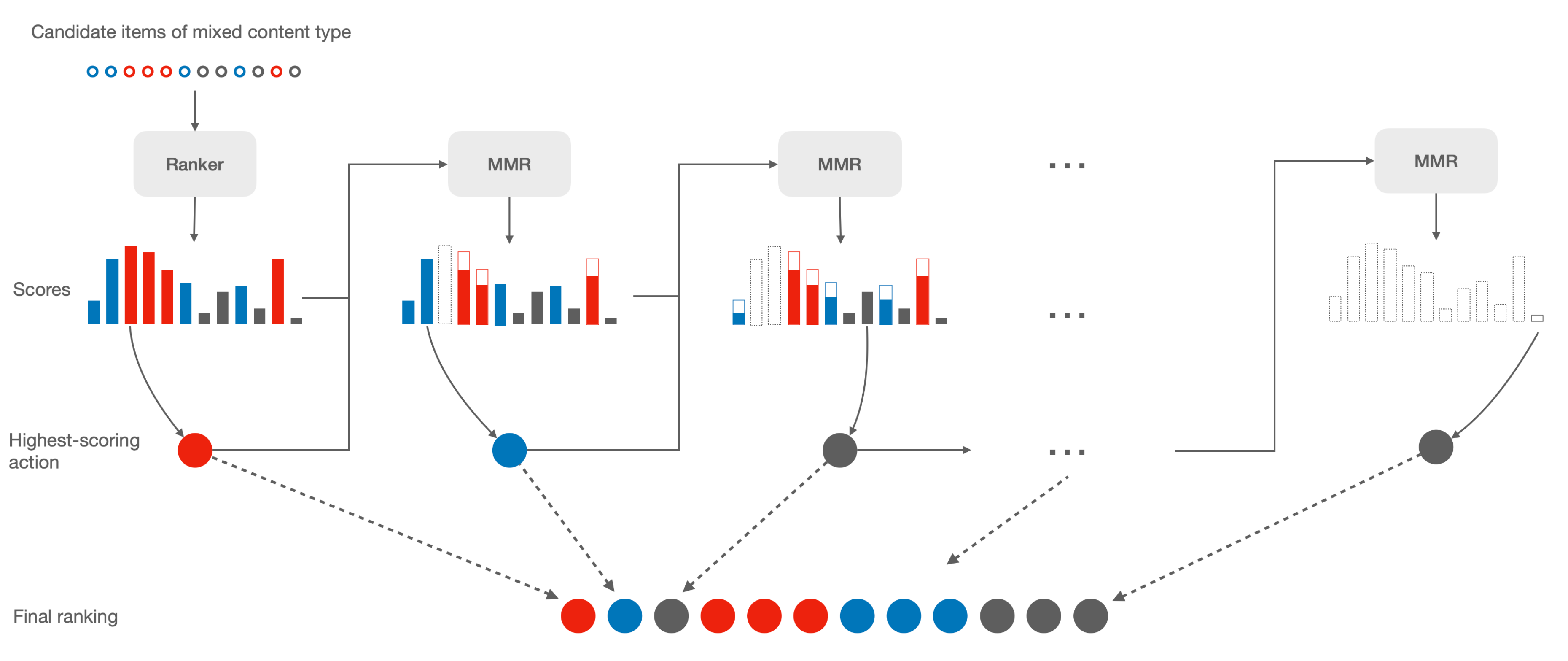}
    \Description[Illustration of MMR.]{A toy illustration of MMR with three content types.}
    \caption{MMR selection for cross-content-type ranking and $C=3$ content types (red, blue, gray). After each action selection, the scores of the remaining candidate items are penalized by their content-type similarity to items that have already been ranked.}
    \label{fig:mmr_illustration}
\end{figure*}

The \textit{Maximal Marginal Relevance} (MMR) re-ranking introduced in ~\citep{carbonell1998use} is a widely used approach for diversification. It requires items relevance scores $s(d_j) = h(x_j)$, a similarity metric $D(d_j, S_k)$ between one item $d_j$ and a list of items $S_k$, and a trade-off parameter $\lambda \in [0, 1]$ to balance relevance and diversity. The re-ranking occurs sequentially: the first ranked item is the one with highest relevance score, and given the first $k$ items $S_k$ in the list, the $(k+1)$-th is chosen, among the remaining items, as the arg max of the score $s'(d_j) = \lambda s(d_j) - (1-\lambda) D(S_k, d_j)$. The similarity score is arbitrary, for instance the maximum of the cosine distance between the candidate item and the already ranked ones based on pre-computed embeddings. In the case presented in the paper where content-types are music and podcast, we opted for a simple similarity metric defined as the proportion of already ranked items from the candidate's content-type, namely $D(d_j, S_k) = \frac{1}{k}\sum_{d\in S_k} \mathbbm{1}[c(d)=c(d_j)]$, penalising items whose content-type has higher exposure in the partial ranking. Figure \ref{fig:mmr_illustration} provides an illustration of a CCT-ranking procedure using MMR.


\subsection{MB with "\textit{at least} $p$\%" exposure guarantees in the $C=2$ case.} \label{app:MB_atleast}
As described in the main text, each parameter $p_c$ of MB maps to the content-type average exposure ($p_c * 100$\% of the slate will be covered by items from content type $c$). Consider a situation with $C=2$ content types, one slow content type $c_s$ and one fast content type $c_f$. Assume that the slow content type would receive, say, 10\% average exposure, and that this average is composed of many users that are even less exposed to items form $c_s$ (because they never engage with items from $c_s$), whereas a few other users are exposed to $c_s$ considerably more than the average. Now assume that for business reasons, the average exposure of $c_s$ should be boosted to 20\%, thus using MB and setting $p_{c_s} = 0.2$. In such a situation, while the majority of users will see a boost in exposure for $c_s$, the few users who had already seen a higher exposure of $c_s$ will see a drop in exposure to 20\%. 
To avoid this case, one can apply the following modification to MB: for a given ranking inference, if the baseline ranker already provides at least the desired exposure to $c_s$, the blending procedure is not triggered and the original ranking is used. This ensures a lower bound on the exposure of $c_s$.

\subsection{MB propensity estimation} \label{MB_props}

A propensity matrix $P \in [0, 1]^{k \times k}$ defines the probability distribution of where in the ranking each candidate item is likely to end up. Propensity matrices are a main building block for many counterfactual LTR algorithms or off-policy evaluators~\cite{li2018offline}. Specifically, $p_{ij} = P[i, j]$ describes the probability of the $i$-th highest-scoring item to end up in position $j$ of the ranking. This probability not only depends on the score of item $i$ but also on (1) the item’s content type (denoted by $c(d_i)$), (2) how many items of the same content type have already been ranked, and (3) how many are yet to be ranked. 

The propensity matrix for MB allows the following closed form solution. If $d_i$ is the $k$-th highest-scoring action of content type $c_i$, then the propensity $P[d_i, j]$ is the probability that $k-1$ actions have been ranked until position $j$ (given by a binomial distribution) times the probability that it will be ranked in position $k$ (given by a Bernoulli trial). Using SciPy~\citep{virtanen2020scipy} notation, this is simply given by \texttt{binom(j-1, $p_{c_i}$).pmf(k-1) * $p_{c_i}$}, where \texttt{binom(n, p).pmf(m)} is the probability that there were \texttt{m} successes in a Binomial trial with \texttt{n} experiments and success probability \texttt{p}.


\end{document}